# Experimental Evaluation of the Effects of Antenna Radiation Characteristics on Heart Rate Monitoring Radar Systems

Mehrdad Nosrati, *Student Member, IEEE*, and Negar Tavassolian, *Member, IEEE*

Stevens Institute of Technology

E-mail: mnosrati@stevens.edu, ntavassolian@stevens.edu

*Abstract*—This paper presents an experimental study to evaluate the effects of antenna radiation parameters on the detection capabilities of a 2.4 GHz Doppler radar used in non-contact heart rate monitoring systems. Four different types of patch antennas and array configurations were implemented on both the transmitter and receiver sides. Extensive experiments using a linear actuator were performed and several interesting and nontrivial results were reported. It is shown that using a linearly polarized single patch antenna at the transmitter and a circularly polarized antenna array at the receiver results in the highest signal quality and system performance. Proof-of-concept experiments on human subjects further validated the suggested results. It was also shown that using the recommended antenna arrangement will boost the heart rate detection accuracy of the radar by an average of 11%.

*Index Terms*—Antenna characterization, circular polarization, Doppler radar, heart rate monitoring, non-contact monitoring.

## I. INTRODUCTION

Noncontact sensing of human vital signals using Doppler radar technology has attracted much attention over the past decades [1]-[6]. There are numerous potential applications for a non-invasive technique to monitor respiration and/or heartbeat. Doppler radar, operating at microwave frequencies in the range of 1-10 GHz, has long been suggested as a valuable tool in intensive care monitoring [1], [4], [7], long-term monitoring [5], [6], and home healthcare applications as well as in non-clinical fields such as health monitoring of field workers (i.e., airplane pilots, firefighters, etc.) [8].

The antenna is an indispensable part of a noncontact monitoring system and strongly affects its performance parameters such as spatial resolution, detection accuracy, and sensitivity [3], [9]-[11]. For instance, an antenna with a wider beamwidth potentially receives more clutter and noise from the environment which could in turn degrade the system performance. It is therefore expected that a system with a directive antenna could provide a higher signal-to-noise ratio (SNR) than an identical system with a wide-beamwidth antenna. Moreover, due to the presence of the human body and body movements, the received signal will suffer from multipath fading. Furthermore, the interaction of the human body with the transmitted waves will lead to a change of polarization in the reflected signal and hence results in polarization mismatch losses and an eventual decrease in the SNR on the receiver side. To date, most studies have been devoted to the development and improvement of transceiver circuits [12]-[14], baseband demodulation and signal processing algorithms [15]-[18], and system architectures [19]-[21]. Significant progress has been achieved in these areas. Most of these works have employed basic linearly-polarized (LP) patch antennas in their systems, for example those reported in [8], [11], [22]-[25]. These simple antennas have a typical gain of about 7 dB and an approximate beamwidth of 80° [26]. Except for [9], [10], [27], [28], no other work has studied the above-mentioned issues and the effects of the employed antennas on their systems. Reference [9] studied four different types of antennas (Yagi-uda, helical, log-periodic, and patch) with a continuous-wave (CW) Doppler radar. However, no combination of these antennas nor with different polarization characteristics was studied. The authors of [10] showed that using an array patch antenna in the transceiver enhances the beat-rate detection accuracy compared to a single patch antenna. However this result was due to the higher gain associated with the array. In [27], the effects of the beamwidth of patch antennas when the subject and radar were misaligned were studied. However, the antennas had different gains, and this affected the accuracy of the results. Recently, the effect of antenna position and its location on the measured signal quality was observed in [33]. Moreover, no discussion was given on the polarization effects of the antennas.

The polarization effect is one the most important factors when designing an optimum radar system. Numerous works have been carried out in this regard in other areas (mainly in outdoor applications) in order to determine the best antenna configuration. For example, vertical-vertical (VV) is the preferred polarization configuration when studying the small-scale roughness of capillary waves on the water surface, or horizontal-horizontal (HH) is the preferred polarization combination in the study of soil moisture because vertically-oriented crops (e.g. wheat and barley) will not have any effects on the EM waves with horizontal polarization, and therefore the backscatter will be mainly dominated by the soil moisture rather than unwanted signals from the crop [29]. However, no such work has been performed in short range indoor healthcare monitoring applications.

Having a detailed knowledge of the effects of the antenna radiation pattern and polarization on the system performance is crucial for developing a robust and reliable system. There is a definite need in the literature for a thorough investigation of these effects. The first intention of this paper is to study the effects of antenna radiation pattern and polarization characteristics in a noncontact monitoring scenario. As such four different types of antennas are studied: 1) a single LP patch antenna, 2) a single circularly-polarized (CP) patch antenna, 3) a 2×2 LP patch array, and 4) a 2×2 CP patch array. There were therefore a set of 16 combinations of these antennas, e.g., a single LP antenna as the transmitting and a CP array as the receiving antennas, or vice versa. The system performance was studied for all combinations. To make accurate and reproducible evaluations, a linear actuator was employed to produce periodic movements. In contrast to [27] and [10], all antennas were designed to have equal gains (i.e., the arrays has the same gains as the single antennas). This ensures that our comparison results are due to antenna radiation and polarization characteristics, not due to a higher SNR level due to higher gain of array antennas. Our results clearly demonstrate that both the transmitting (Tx) and receiving (Rx) antennas play an important role and have significant impacts on the detection capabilities of the radar system. After rigorous time and frequency domain analyses, the best case among the 16 combinations is identified and the results are explained and justified. Several guidelines are also given for selecting an antenna with appropriate polarization and beamwidth characteristics in a radar detection setup. Finally, proof-of-concept human experiments were performed to further evaluate the results and highlight the importance of the work.

This paper is organized as follows. In Section II, the system hardware including the radar sensor and antennas are described. The experimental setup is presented in Section III. Section IV covers the measurement results and discussions. A pilot study on human subjects is presented in Section V. Conclusions are summarized in Section VI.

## II. SYSTEM HARDWARE

The hardware used in the system is described in this section. As shown in Fig.1, a bistatic quadrature Doppler radar emits a single-tone signal through the transmitter antenna. A linear programmable actuator (PA-14 from progressive automations ) is used as the moving target. The reflected signal from the target is captured by the receiving antenna, demodulated, amplified, sampled by an analog-to-digital (A/D) converter, and fed to a laptop for processing.

### A. Doppler Radar Sensor

Based on Doppler radar theory, the displacement of the target will result in a phase shift in the reflected signal. By properly demodulating this phase information, the displacement information can be obtained. However, higher-order harmonics of the desired signal are also produced after demodulation [3], [20]. Single-channel Doppler radars suffer from a well-known problem called null detection [10], [20], [21], [24], [30]. This problem results in the fading of odd-order frequencies (including the desired fundamental frequency and accentuation of even-order harmonics. Using a quadrature Doppler radar transceiver alleviates this issue [4], [24], [25]. Fig. 1 shows the quadrature radar system employed in this work and its functional block diagram. The radar circuit was obtained from yearONE LLC. The output power level of the radar is 0 dBm. A direct-conversion receiver architecture is used for recording the subject movement. As shown in Fig. 1, the voltage-controlled oscillator (VCO) in the transmitter chain generates a carrier signal at 2.4 GHz. The VCO signal passes through a power splitter and is then transmitted from the Tx antenna. The other half of the VCO signal is injected to the mixer in the receiver chain. The reflected signal is captured by the Rx antenna and amplified by a low-noise amplifier (LNA). A band-pass filter is adopted to remove out-of-band interferences. After down-conversion of the received signal, a baseband amplifier is used to amplify the baseband signals. The transmitted signal $T(t)$ can be expressed as:

$$T(t) = A\cos(2\pi f t + \phi(t)) \tag{1}$$

where $A$ is the amplitude of the signal, $f$ is its frequency, and $\phi(t)$ is its phase. Denoting the displacement of the target by $x(t)$, the received signal $R_m(t)$ can be expressed as:

$$R_m(t) = KA\cos\left(2\pi f t - \frac{4\pi d_0}{\lambda} - \frac{4\pi x(t)}{\lambda} + \phi(t - \frac{2d_0}{c})\right) \tag{2}$$

where $K$ is the reduction of the amplitude due to path loss and reflections from the subject, $d_0$ is the distance between the system and the subject surface, $\lambda$ is the wavelength of the radar carrier wave, and $\theta_0$ is the phase shift due to reflections from surfaces. After down-conversion of the received signal $R_m(t)$, two baseband signals, namely the in-phase signal $R_I(t)$ and the quadrature signal $R_Q(t)$ with a phase difference of $\pi/2$ will be generated:

$$R_I(t) = A_I \cos(\frac{4\pi x(t)}{\lambda} + \theta_0 + \Delta\phi) \tag{3}$$

$$R_Q(t) = A_Q \sin\left(\frac{4\pi x(t)}{\lambda} + \theta_0 + \Delta\phi\right) \quad (4)$$

where $A_1$ and $A_Q$ are the amplitudes of the in-phase and quadrature signals and $\Delta\phi$ is the total residual phase noise. Assuming that the periodic movement signal $x(t)$ in (1) and (2) can be approximated as $x(t) = k\, Sin(\omega t)$, we can write either of the $I$ and $Q$ channels (for example the $I$ channel) as:

$$R_I(t) = \text{Re}\left(e^{j\left(\frac{4\pi k \sin(\omega t)}{\lambda} + \varphi\right)} \times e^{j(\theta_0 + \Delta\phi)}\right) \quad (5)$$

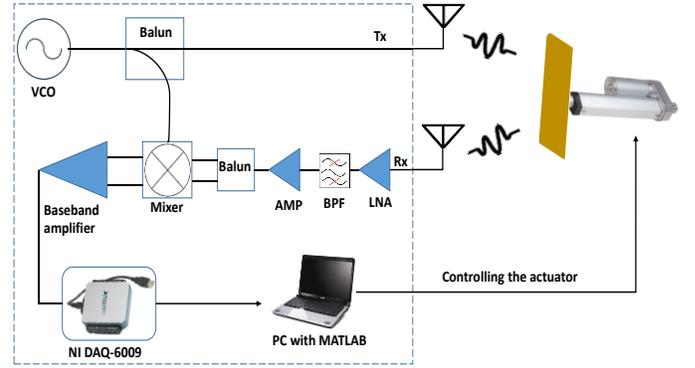

Fig. 1. Overall system configuration. A block diagram of the Doppler radar is shown. A linear actuator with an attached metal sheet is used as the moving subject.

Also, from Fourier transform theory we know that:

$$e^{j\left(\frac{4\pi k \sin(\omega t)}{\lambda}\right)} = \sum_{n=-\infty}^{\infty} J_n\left(\frac{4\pi k}{\lambda}\right) \times e^{jn\omega t} \quad (6)$$

where $J_n(x)$ is the $n^{th}$ order Bessel function of the first kind. Combining (5) and (6):

$$R_I = \sum_{n=-\infty}^{\infty} J_n\left(\frac{4\pi k}{\lambda}\right) \times \cos(n\omega t + \theta_0 + \Delta\phi) \quad (7)$$

With some manipulations and considering $\varphi = \Delta\phi + \theta_0$ we can rewrite (7) as:

$$R_I = -2\sum_{n=1}^{\infty} J_{2n}\left(\frac{4\pi k}{\lambda}\right) \times \cos(2n\omega t) \times \cos(\varphi)$$
$$- 2\sum_{n=0}^{\infty} J_{2n+1}\left(\frac{4\pi k}{\lambda}\right) \times \sin\big((2n+1)\omega t\big) \times \sin(\varphi) \quad (8)$$

Thus, the frequency spectrum of a pure sinusoidal movement signal will contain the fundamental frequency and other higher order harmonics, where the amplitude of each harmonic is proportional to a modified Bessel function. As can be seen from this equation, if $\varphi = 0°$, then the amplitude of the fundamental frequency goes to zero and the second harmonic becomes the dominant harmonic. The complex signal demodulation [24] can eliminate the above null detection problem by combining the $R_I(t)$ and $R_Q(t)$ signals in the baseband. The complex signal $R(t)$ can be formed by:

$$R(t) = R_I(t) + j.R_Q(t) = e^{j\left(\frac{4\pi x(t)}{\lambda} + \varphi\right)} \quad (9)$$

By following the same procedure mentioned above and applying (6) to (9) we get:

$$R(t) = R_I(t) + j.R_Q(t) = \sum_{n=-\infty}^{\infty} J_n\left(\frac{4\pi k}{\lambda}\right) \times e^{jn\omega t} \times e^{j\varphi} \quad (10)$$

In this equation, the harmonic amplitudes do not rely on $\varphi$ and the desired fundamental frequency component will always be present in the spectrum.

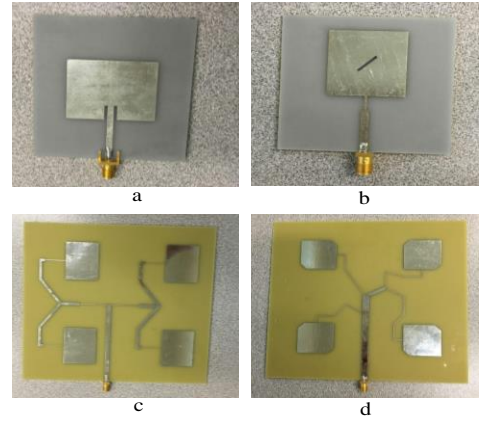

Fig. 2. Photographs of the fabricated antennas. (a) LP single patch antenna, (b) CP single patch antenna, (c) LP array, and (d) CP array.

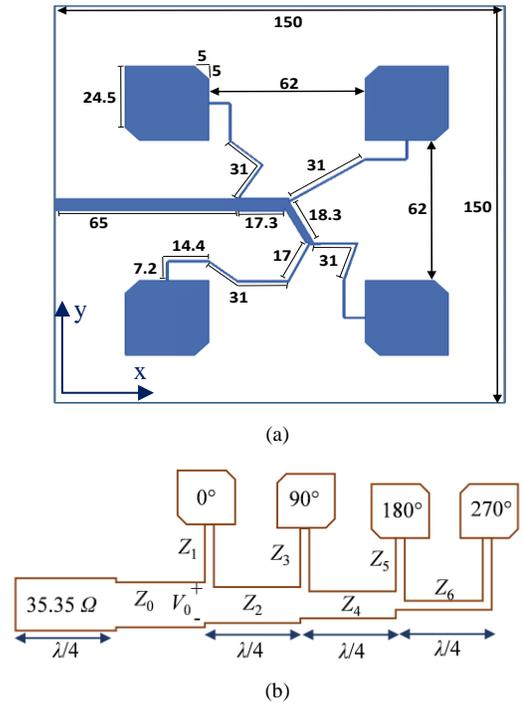

Fig. 3. (a) The geometry and dimensions (in mm) of the designed 2×2 CP array antenna. (b) A simplified schematic of the feeding network.

### B. Antenna Design

In this paper, the effects of antenna polarization and beamwidth on non-contact monitoring radar systems is examined. Four different types of the most commonly-used microstrip patch antennas are designed and fabricated. The fabricated antennas are as follows: a single LP patch antenna, a single CP patch antenna, a 2×2 LP patch array, and a 2×2 CP patch array. Fig. 2 shows photographs of the fabricated antennas. Since the Doppler radar used is bistatic, two of each antenna type have been fabricated, one for the Tx and the other for the Rx. The design process for the antennas, except for the 2×2 CP array, is relatively simple and straightforward.

#### 1) 2×2 CP Antenna Array Design

Designing the CP array requires careful considerations as will be presented below. The geometrical schematic of the proposed 2×2 CP array is shown in Fig. 3 (a). There are many designs for the single-feed patch that generate circular polarizations [31]. However, the array elements here are four truncated-corner patch antennas as shown in Fig. 3 (a). The corresponding feeding network is also shown in this figure. Simulation results indicate that the single-fed truncated-corner square patch antenna has $|S_{11}| < -10$ dB and $|AR| < 3$ dB bandwidths of 10% and 5% at 2.4 GHz, respectively. The four patches are placed in a planar arrangement.

The array geometry is designed using the sequential-rotation technique [32]. As seen from Fig. 3 (a), the distance between adjacent array elements is $\lambda/2$ ($\lambda$ is the free space wavelength) which These patches receive the signal from the feed-line with 0°, 90°, 180°, and 270° phase differences respectively. Designing the feeding network is another challenging part of the design since it needs to deliver an equal amount of power to each patch antenna and also provide the necessary phase differences to generate a circularly-polarized wave. Fig. 3 (b) demonstrates the array feeding network. T-junction power dividers are incorporated into the feed-line. Following the notations of Fig. 3 (b) and assuming an equal amount of power for each array element, we have:

$$P_{in} = \frac{V_0^2}{2Z_0} \qquad (11)$$

$$P_1 = \frac{V_0^2}{2Z_1} = \frac{1}{4}P_{in} \Rightarrow Z_1 = 4Z_0, \quad P_2 = \frac{V_0^2}{2Z_2} = \frac{3}{4}P_{in} \Rightarrow Z_2 = \frac{4}{3}Z_0$$

$$P_3 = \frac{1}{3}P_2 \Rightarrow Z_3 = 3Z_2 = 4Z_0, \quad P_4 = \frac{2}{3}P_2 \Rightarrow Z_4 = \frac{3}{2}Z_2 = 2Z_0$$

$$P_5 = \frac{1}{2}P_4 \Rightarrow Z_5 = 2Z_4 = 4Z_0, \quad P_6 = \frac{1}{2}P_4 \Rightarrow Z_6 = 2Z_4 = 4Z_0$$

By choosing $Z_0 = 25\ \Omega$ and using (11), we can calculate the impedance values of the lines. For the calculated characteristic impedances $Z_0$ to $Z_6$ and dielectric constant $\varepsilon_r$, the line width $W$ can be found from the following formula for microstrip lines [33]:

$$W = d.\left(\frac{8e^A}{e^{2A}-2}\right)$$
$$A = \frac{Z_0}{60}\sqrt{\frac{\varepsilon_r+1}{2}} + \frac{\varepsilon_r-1}{\varepsilon_r+1}\left(0.23+\frac{0.11}{\varepsilon_r}\right) \qquad (12)$$

In this equation $d$ is the substrate height. As shown in Fig. 3 (b), a 35.35 $\Omega$ quarter-line transformer is exploited to transform the $Z_0 = 25\ \Omega$ feed-line into 50 $\Omega$.

### 2) Antennas Fabrication and Measurement Results

The LP and CP array antennas were fabricated on FR-4 substrates with $\varepsilon_r = 4.4$ and $tan\delta = 0.017$ and thickness of 1.6 mm. The single LP and CP antennas were printed on low-loss RT-Duroid 5880 substrates with $\varepsilon_r = 2.2$ and $tan\delta = 0.0007$ and thickness of 1.6 mm. This eliminates the gain differences between the single and array antennas resulting in an accurate evaluation of the antenna beamwidth effects on the radar performance. The fabricated antennas were characterized in an anechoic chamber. The $S_{11}$ and AR results are presented in

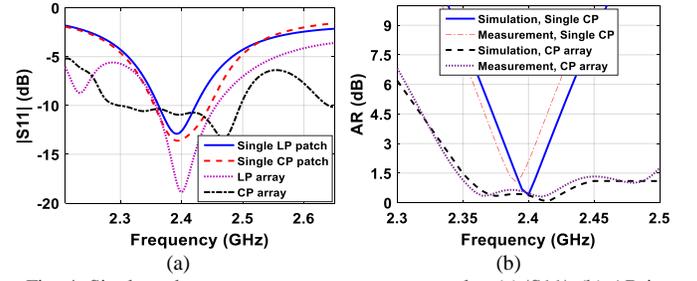

Fig. 4. Single and array antenna measurement results. (a) $|S_{11}|$, (b) AR in the boresight direction ($\varphi = \theta = 0°$).

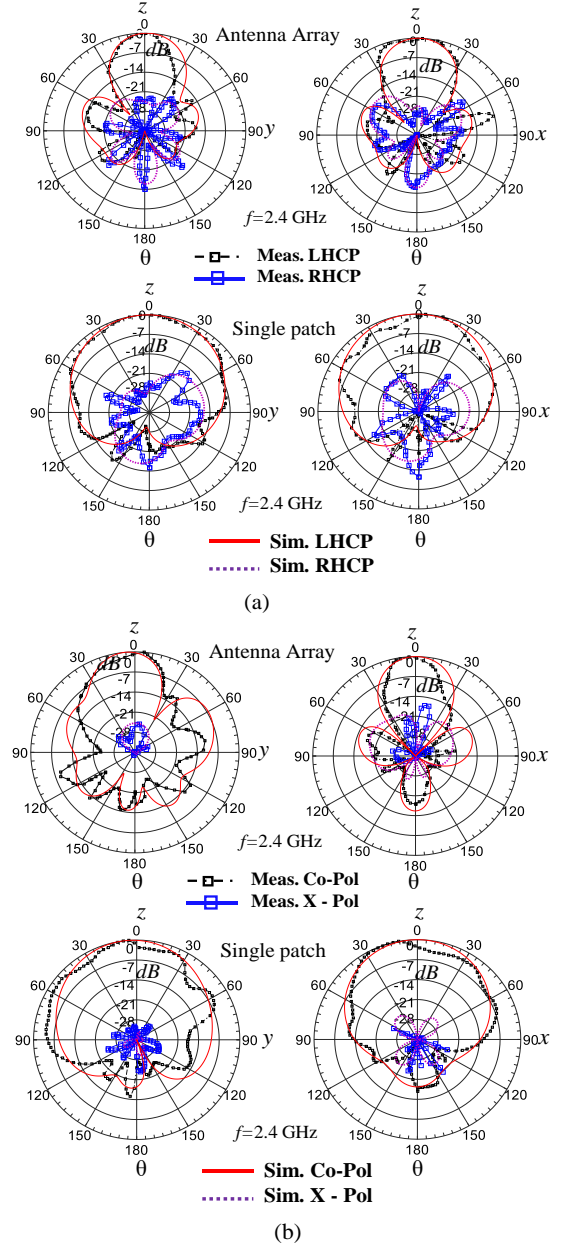

Fig. 5. Normalized radiation patterns of (a) CP single and array antennas, and (b) LP single and array antennas for $\varphi = 0°$ and 90° versus $\theta$.

Figs. 4 (a) and (b), respectively. The normalized $E$-plane radiation patterns are depicted in Fig. 5. The measured peak gains of all antennas are within $5.8 \pm 0.3$ dBi. The half-power beamwidth (HPBW) is 37° for the array antennas and 81° for the single antennas. Note that a left-handed CP wave after reflection from the target metal sheet will become a right-handed CP wave and cannot be received by the same antenna polarization. Therefore, we fabricated both left and right handed CP antennas for the cases where both Tx and Rx configurations require a CP antenna.

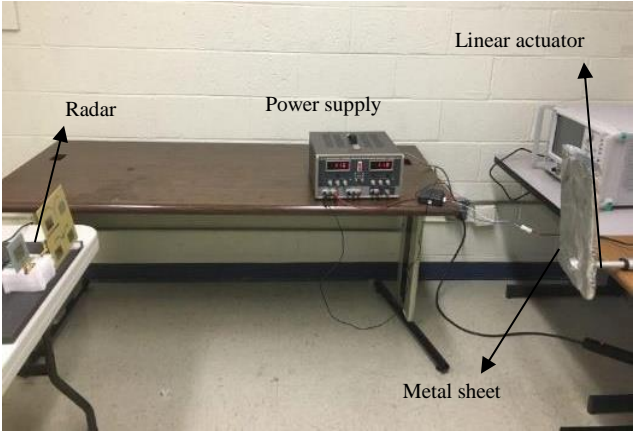

(a)

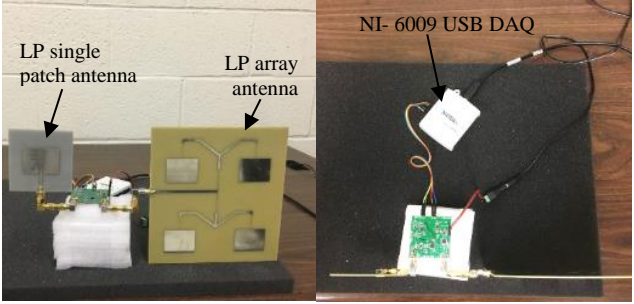

(b)                                    (c)

Fig. 6. (a) The measurement setup. (b) A photo of the radar with a CP array antenna on the Tx side and an LP single antenna on the Rx side. (c) Top view of the radar circuit.

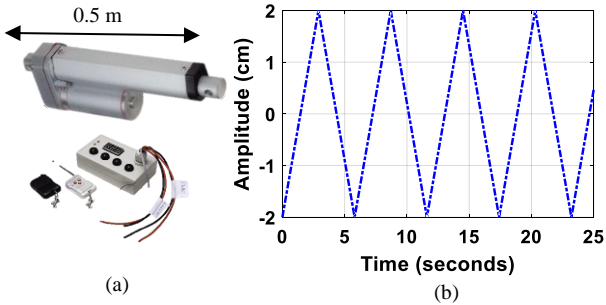

(a)                                    (b)

Fig. 7. (a) The actuator with its controller box. (b) The actuator movement pattern. The actuator movement has a period of 5.8 seconds and an amplitude of 2 cm.

## III. TEST SETUP AND MEASUREMENTS

Experiments were performed in a laboratory environment. Fig. 6 illustrates the measurement setup. As already mentioned above, a linear actuator has been employed to maintain a constant movement pattern. Testing the radar on human subjects would not be suitable for comparison purposes as the human breathing and heart rates are not consistent with each other. Furthermore, the subject movement can easily disturb the measured signal [3], [22], [24]. The linear actuator along with the power supply and the Doppler radar are seen in Fig. 6 (a). A 30 cm × 20 cm plate with an aluminum foil covering it is attached to the actuator. The aluminum foil reflects the major portion of the incident wave energy from the radar. The actuator and radar are placed on separate tables with the center of the metal plate facing the radar. The distance between the metal plate and the radar is 1.5 m. The radar setup for the case where the Tx antenna is a LP array and the Rx antenna is an LP single patch is demonstrated in Fig. 6 (b). Figs 7 (a) and (b) show the actuator and its movement signal, respectively. The linear actuator produces a periodic back-and-forth movement with a constant period of 5.8 seconds and an amplitude of 2 cm. The duration of each experiment is 30 seconds. The baseband signal received from the Doppler radar is sampled at 100 Hz and fed to the laptop using a NI data acquisition (DAQ) device (NI USB 6009) . After DC offset cancellation (subtracting the signal average from the raw signal), we calculate the frequency spectrum of the recorded data as follows:

$$S(f) = \left| FFT(I + jQ) \right| \tag{13}$$

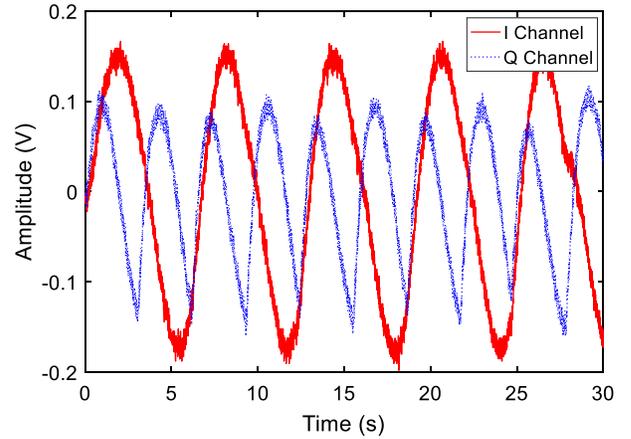

Fig. 8. A sample of the measured data in time domain. The $Q$ channel has twice the frequency of the $I$ channel, indicating it is in a null point.

## IV. EXPERIMENTAL RESULTS AND DISCUSSION

### A. Experimental results

Fig. 8 shows a sample of the measured signals in time domain. It can be seen that the $Q$ channel has twice the frequency of the $I$ channel. Also, the $Q$ channel has a smaller amplitude which indicates that the $Q$ channel is in the null mode and therefore even harmonics of the fundamental frequency are emphasized [24], [25], [30]. The goal of non-contact monitoring applications using a Doppler radar is to extract and recover useful information from a moving subject, mainly its movement frequency and amplitude. In the case of detecting human heartbeat, the strength of the fundamental frequency in the frequency spectrum of the received signal is generally considered as the most important performance metric [6], [22], [24]. A stronger fundamental frequency results in a higher SNR in the receiver and makes it simpler to estimate the movement frequency and extends the detection range of the radar. The frequency spectrums of the received signals for the 16 different antenna configurations tested are shown in Fig. 9. This figure is plotted as a 4×4 matrix, with each element representing one of the 16 combination sets of antennas. For instance, the element at the second row and third column shows the case where the Tx antenna is an LP array and the Rx antenna is a CP single patch. Please note that these plots show raw data without any filtering or pre-processing steps. To ensure the accuracy of results, each experiment was repeated 5 times and the signals were averaged in time domain to yield the final signal employed. The amplitudes of the first two harmonics corresponding to the *I* and *Q* channels respectively are marked with red arrows on each figure along with their values in dB. According to Fig. 9, when Tx uses an LP single patch and Rx employs a CP array (corresponding to the fourth row and first column), the received signal has the strongest amplitude at the fundamental frequency and therefore shows the best performance. The fundamental frequency amplitude is -6.41 dB in this case. On the other hand, the plot in the second row and fourth column with the CP array on the Tx side and the CP single patch on the Rx side shows very poor performance (fundamental frequency amplitude of -22.77 dB).

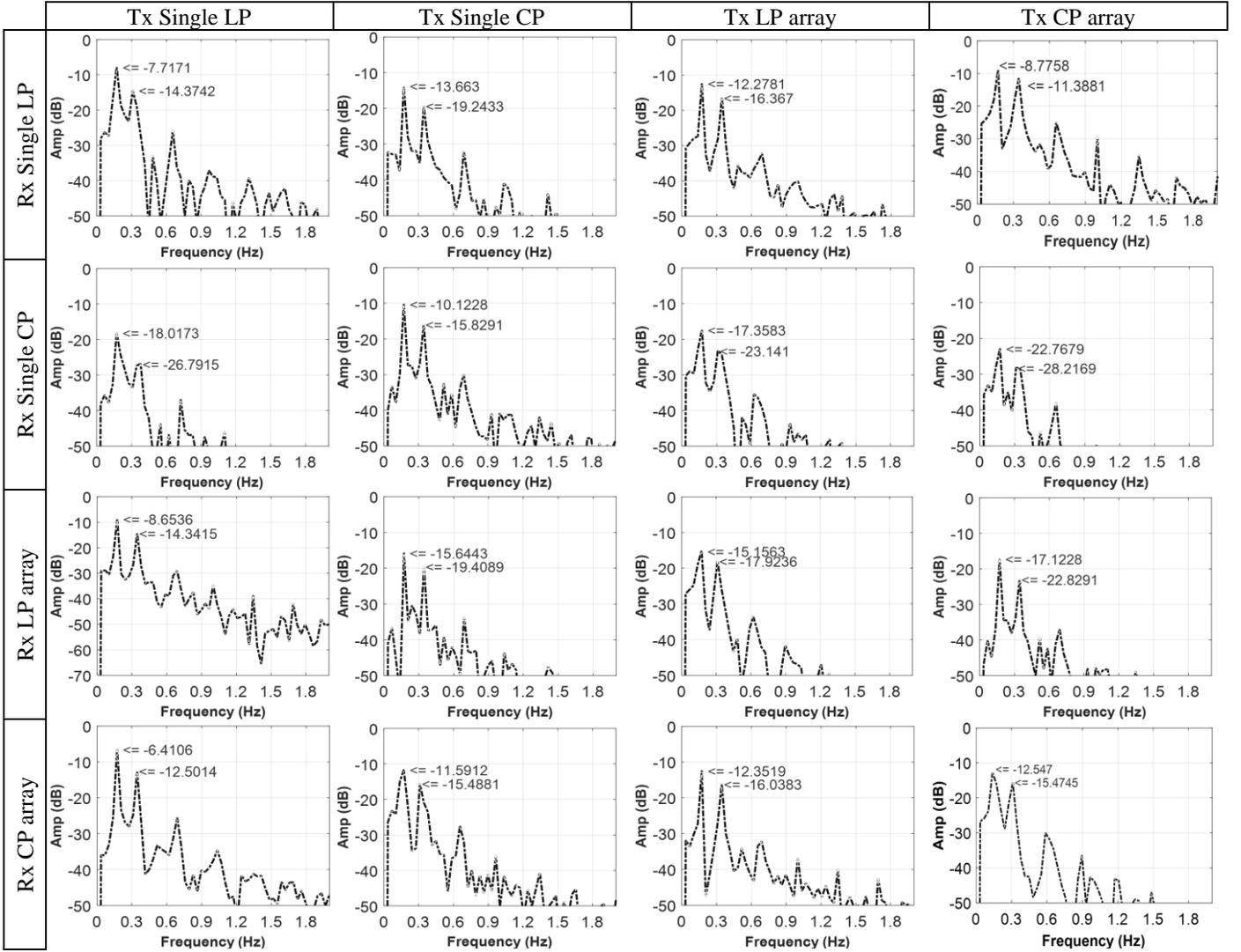

Fig. 9. Frequency spectrum of the different measurement scenarios. The figure is plotted as a 4×4 matrix, which each element of the matrix representing one of the 16 combination sets.

### B. Discussion

Fig. 9 reveals several interesting and non-trivial results. For instance, it may be assumed that swapping the Tx and Rx antennas would not affect the received signal significantly. However, our measurements indicated that this assumption is not valid for short-range indoor human vital signs detection. For example, consider the case where an LP array and an LP single patch are used for the Tx and Rx antennas respectively, i.e., the first row and third column in Fig. 9. The peak value of this plot is -12.27 dB, while swapping the antennas (corresponding to the third row and first column) changes the peak value to -8.56 dB. In other words, our radar antenna configuration is not reciprocal. This result was validated and verified by several follow-up experiments. This non-reciprocity behavior is attributed to two major phenomena: 1) different radar cross sections for each set of the antenna arrangements, and 2) presence of multipath reflections which are different for each case. In the following paragraphs, the effects of each of these phenomena will be elaborated and discussed.

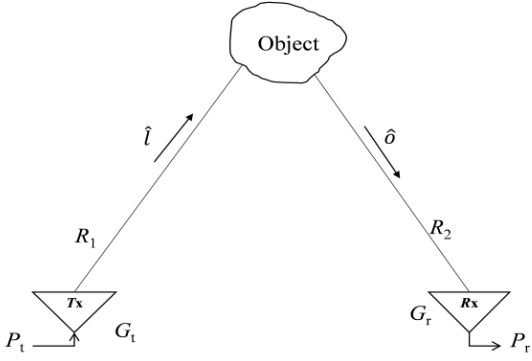

Fig. 10. Radar geometry with the notations of (14).

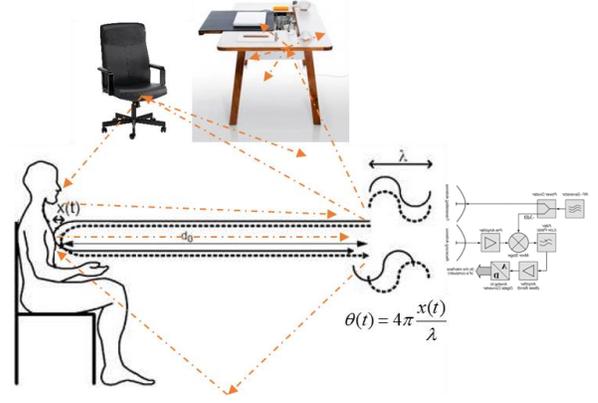

Fig. 11. A simplified model for explaining the ray-tracing effect in our scenario.

### 1) The Effects of Radar Cross Section

To investigate the effects of the radar cross section on its performance, it should be considered that the radar cross section is not only dependent on the target features such as its geometry, material, and orientation, but is also directly affected by the incident wave characteristics and the arrangement of the Tx and Rx antennas. In a general setup of a bi-static radar, the received power $P_r$ can be calculated using the following equation [34]:

$$P_r = P_t G_t(\hat{i}) G_r(-\hat{o}) \left( \frac{\sigma(\hat{o},\hat{i}) \lambda^2 \rho(\hat{o},\hat{i})}{(4\pi)^3 R_1^2 R_2^2} \right) \quad (14)$$

In this equation, $\hat{i}$ and $\hat{o}$ are unit vectors in the directions from the Tx antenna to the object and from the object to the Rx antenna, respectively (Fig. 10). $G_t(\hat{i})$ and $G_r(-\hat{o})$ are the gains of the Tx and Rx antennas, which are functions of the antenna beamwidth, polarization, and realized gain. Pt is the transmitted power and $R_1$ and $R_2$ account for the distances from the target to the Tx and Rx antennas, respectively. $\lambda$ is the wavelength and $\rho(\hat{o},\hat{i})$ is the polarization mismatch factor between the transmitted and reflected wave polarizations with its value limited to the range of [0,1] where 0 indicates a complete mismatch. $\sigma(\hat{o},\hat{i})$ is the radar cross section given in (15):

$$\sigma = 4\pi\sigma_d(-\hat{i},\hat{o}), \quad and \quad \sigma_d(\hat{i},\hat{o}) = |f(\hat{i},\hat{o})|^2 \quad (15)$$

$f(\hat{i},\hat{o})$ is named the scattering amplitude and represents the amplitude, phase, and polarization of the scattered wave in the far field in the direction of $\hat{o}$ when the object is illuminated by a plane wave propagating in the direction $\hat{i}$. It should be noted that even if the incident wave is linearly polarized, the scattered wave is in general elliptically polarized [34]. As can be seen from (14) and (15), the radar cross section of the object, and as a result the received signal level, is directly affected by the characteristics of the transmitter and receiver antennas. As such, the beamwidth of the antennas and their polarization characteristics (which are included in $G_t(\hat{i})$ and

$G_r(-\hat{o})$ ) have significant effects on the radar cross section ( $f(\hat{i},\hat{o})$ ). It is clear that for each set of the Tx and Rx antennas, the gain functions, polarization mismatch factor, and the scattering amplitude function are different, leading to different received signal levels. Also, we should mention that the radar cross section of either a very large or a very small object at far distances approaches a constant value (proportional to its geometric cross section) [34]. This is the main reason that in the outdoor radar scenario, the non-reciprocity effect is less apparent and severe (in outdoor radars, objects are often far away from the radar).

### 2) Multipath Reflection Effect

Another phenomenon that contributes to the radar non-reciprocity is the multipath effect. Multipath fading is a well-known phenomenon in wireless communication systems as it can easily wipe out the desired signal due to out of phase combinations at the receiver side [35]. Despite the fact that in our case multipath effects are not as strong as the radar cross section effects, they should still be considered. We employ the Ray Tracing Technique [36] as a tool to explain and investigate the multipath reflection contribution on the non-reciprocal behavior of the system. In our indoor scenario, there are various objects present which have comparable dimensions to the wavelength. A simple example is shown Fig. 11 where in addition to the human body (or any other objects such as the actuator), the walls, the ceiling, and the floor, there are also a chair and a table present in the test room. The Rayleigh criterion define the critical height of surface protuberance ($h_c$) for a given angle of incidence ($\theta_i$) as [37]:

$$h_c = \frac{\lambda}{8.\sin(\theta_i)} \quad (16)$$

where $\lambda$ is the wavelength. A surface is considered smooth if its minimum to maximum protuberance h is less than hc, and rough if the protuberance is greater than hc. At the frequency of 2.4 GHz and for a perpendicular incident wave, hc is calculated using (16) as 1.56 cm. Therefore, it is plausible to assume that the human chest (with clothing) is a rough surface since in practice the dents and wrinkles of clothing are approximately 1-2 cm long. The floor, chair, and table are smooth reflectors and the reflection from the ceiling is neglected due to it being far away from the experiment. Using the Ray Tracing method, at the Rx side we will receive: 1) a direct reflection from the object, 2) several second-order reflections from the floor, the ceiling, the chair, and the table, each with a different polarization type (LP or CP) and sense (H, V, RHCP or LHCP), and 3) several higher order (albeit smaller in amplitude) reflections from the scattering objects, again each with a different polarization type and sense. Now, if we swap the antennas, these direct and indirect Tx-to-Rx signals will experience different changes in their polarization characteristics depending on the Tx polarization and the order of the objects they encounter. This causes the received signal to be different in each case and results in the observed non-reciprocity phenomenon.

### C. Effects of Coupling and Noise Floor

Another interesting observation can be made when comparing single antenna patches with array elements. It is expected that when array antennas with narrower beamwidths are exploited in Tx and Rx sides, the reception of unwanted signals would decrease, resulting in a higher SNR level. However, measurements indicate that this conclusion is not valid. For example, consider the third row and third column in Fig. 9. In this case, both Tx and Rx use LP array antennas. The peak value of this plot is -15.16 dB which is 7.44 dB lower than the case where both Tx and Rx sides have LP single patch antennas (the first row and first column with a peak value of-7.72 dB). The same result is observable when comparing the case where both Tx and Rx sides have CP arrays (the forth row and forth column with a peak power of - 17.13 dB) versus the case where they both have CP single patches (second row and second column with a peak power of -10.13 dB). Furthermore, by comparing the third column with the third row or the fourth column with the fourth row, it is observed that employing an array antenna with a narrower beamwidth on the receiver side results in a higher peak power compared to using the same array on the transmitter side. For instance, consider the case in the fourth row and second column. In this case, Rx and Tx antennas are a CP array and a CP single patch, respectively and the peak power is -11.59 dB. If the antennas are swapped (second row and forth column) the peak power becomes -22.77 dB which is significantly lower than the previous case. The same trend is true for other combinations. So, when considering the use of an array on either the Tx or Rx sides, it is advisable to place it on the Rx side. Antenna polarization is also shown to have a great impact on the quality of the received signal although its impact is not as consistent. Still, by comparing the third and fourth rows in Fig. 9 we conclude that using a CP array in the Rx always results in a stronger signal compared to using an LP array in the Rx. Please note that in real-case applications, the depolarization effect due to the human body is much more severe than the aluminum foil and will further increase polarization mismatch losses [37, 38]. Therefore, the polarization impact will become much more discernible.

Regarding the enhanced performance of the single LP-CP array arrangement compared to the general case of single LP-single LP configuration, we believe there are two main reasons for this effect. First, in a general radar setup such as the one discussed in our paper, the Tx and Rx antennas are placed in close proximities (generally within 4-8 cm of each other). Therefore the direct signal transmitted from the Tx antenna to the Rx antenna is comparable in strength to the desired reflected signal from the target. As such, even a small portion of signal leakage from Tx to Rx can desensitize the receiver chain and destroy the desired target signal. Furthermore, this signal leakage will cause a DC offset and degrades the A/D performance. The isolation between the Tx and Rx antennas is especially vital in the case of monitoring human subjects as the reflected signal from the heart-beat is extremely small and can be easily buried in the background noise and interferences. It is generally understood that for two isotropic antennas communicating in free space, the polarization mismatch between an LP-CP arrangement is 3 dB higher than that of an LP-LP arrangement [26]. In this work, we are observing a similar effect in our radar setup. We are also using this effect to our benefit to improve the Tx/Rx isolation and reduce unwanted signal leakage from Tx to Rx. Accordingly, the signal leakage in the system can be reduced by: 1) using an LP-CP combination (rather than LP-LP), and 2) using an antenna array on one side, since an array has a narrower beamwidth than a single patch and therefore reduces interferences. In order to better understand the coupling effect, we measured the S21 of the 16 combination sets of antennas using an Anritsu MS4647B 70-GHz vector network analyzer (VNA). The measurements were performed in the same environment where the experiments took place. Fig. 12 shows the measurement setup. The antennas were placed side-by-side, 6 cm apart from each other.

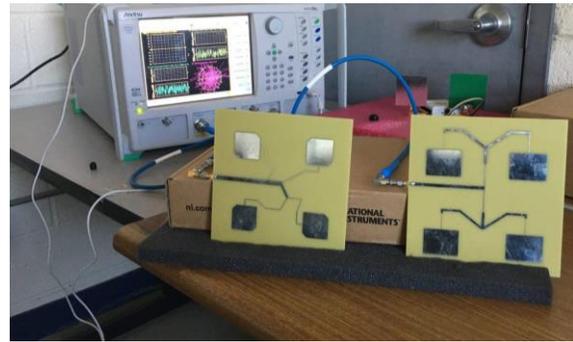

Fig. 12. The measurement setup of |S21|. The figure shows the coupling between LP and CP array antennas.

The measurement results in Fig. 13 follow the expected trends. The |S21| plot shows the coupling between two antennas, therefore the lower this value the higher the isolation (which is desired). The isolation between two identical configurations (i.e. single or array) is better when one of the antennas is CP and the other one is LP. For instance, the isolation between a single LP and a CP array (first row and fourth column) is 36.29 dB which is 4.74 dB better than a single LP and a LP array (first row and third column). Also, when an antenna array is used, the coupling is further reduced. For instance, the isolation when Rx is single CP and Tx is single LP (second row and first column) is 30.66 dB while if we use a LP array in the Tx (second row and third column) this value increases to 34.56 dB. Please note that in these measurements, the S21 is measured between a pair of antennas facing free space and as a result S21=S12. Also, although using an LP array and a CP array (third row and fourth column

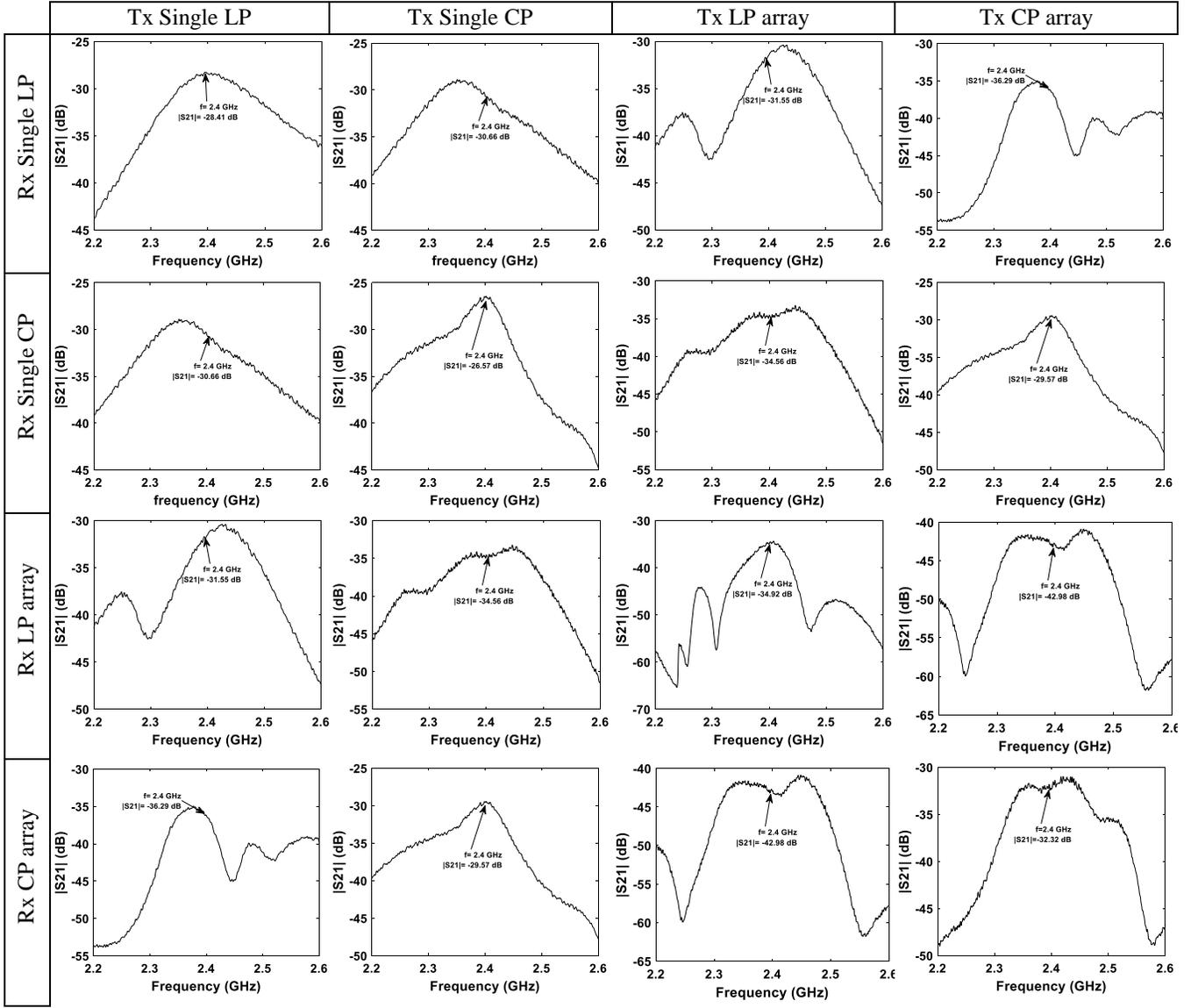

Fig. 13. Measured |S21| of the sixteen different antenna combination sets. The figure is plotted as a 4×4 matrix, which each element of the matrix representing one of the 16 combination sets.

or fourth row and third column) provides the highest isolation, it does not provide the best case in Fig. 9. The reason is that a single patch antenna has a wider beamwidth and can illuminate a larger area of the target which increases the backscattered signal level. Our measurements (Fig. 9 and Fig. 13) confirm that this effect is more influential.

The second reason for the enhanced performance of the single LP-CP array arrangement is that when the transmitted wave hits the moving target, a portion of the wave-front will be reflected with a Doppler shift and a polarization change. This change in polarization depends on the distance between the radar and the target, the angle of incidence, the target's surface area, and the roughness of the target's surface [38]. A circularly-polarized receiver antenna will be able to capture both horizontally-polarized and vertically-polarized incoming waves. On the other hand, a horizontal LP antenna cannot receive any incoming signal from a vertical LP antenna or vice versa. Therefore, by using a CP antenna in the receiver we will ensure that no matter how the Tx polarization changes, a portion of the incoming wave will be received. This is also one of the main reasons that in practice, mobile and WiMAX base stations exploit CP antennas or dual-polarized antennas. Again in this paper, we have used this effect to our benefit by using a CP antenna at the receiver side.

Another interesting phenomenon which was observed during measurements was that the proposed "single LP-CP array" arrangement has a lower noise floor compared to the "single LP-single LP" arrangement. Fig. 14 shows the system noise floor (i.e., the frequency response of the radar while aimed at free space) for these two cases. As seen in Fig. 14 (a), the system noise floor for the "single LP-single LP" case is almost -75 dBm which is 10 dB higher than the noise floor when Tx is a LP and Rx is a CP array antenna (Fig. 14 (b)). The reason for this is that the LP-CP case receives much lower interference and clutter signals. Also, the energy coupling and signal leakage from the Tx to Rx in this case are lower than the LP-LP case. The effect of the noise floor shows itself more clearly if the received data is filtered before the FFT operation.

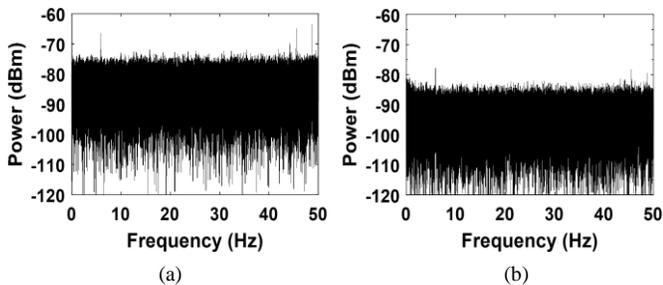

Fig. 14. The radar noise floor. (a) When both Tx and Rx are LP single patch antennas. (b) When Tx is an LP single patch and Rx is a CP array. The latter has a lower noise floor.

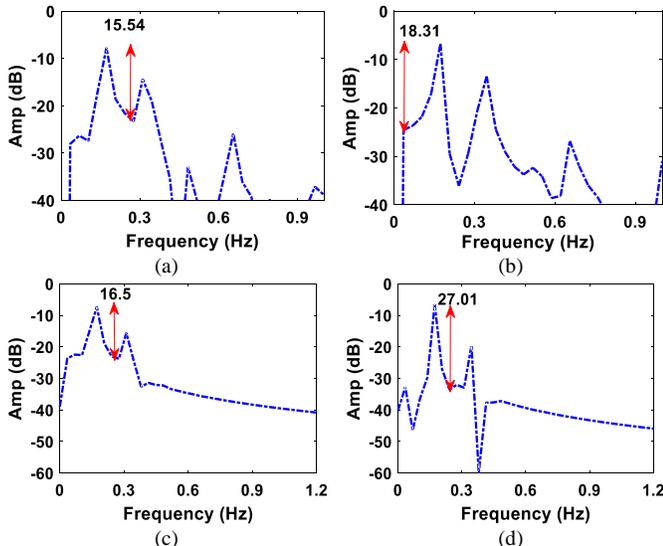

Fig. 15. Frequency spectra of the two discussed cases before and after applying a fifth order Butterworth low-pass filter. (a) "Single LP-single LP" before filtering, (b) "single LP-CP array" before filtering, (c) "single LP-single LP" after filtering, (d) "single LP-CP array" after filtering.

To demonstrate this, we first filtered the data with a fifth order Butterworth low-pass filter with a cut-off frequency of 0.35 Hz (slightly higher than the second harmonic of the actuator) and then calculated the frequency responses. Fig. 15 depicts the frequency spectrums for the two cases discussed above. The upper row corresponds to the spectrums before filtering and the lower row to the spectrums after filtering. Before filtering, the differences between the signal peak level and the noise level, SNR, are 15.54 dB and 18.31 dB for the "single LP-single LP" and "single LP-CP array" cases, respectively (Figs 15 (a) and (b)). These values increase to 16.5 dB and 27.01 dB after filtering (Figs 15 (c) and (d)). Therefore, the SNR improvements due to filtering are 0.96 dB and 8.70 dB for LP-LP and LP-CP cases respectively.

## V. HUMAN SUBJECT STUDY

In this section, a pilot study on human subjects is presented. To validate the results achieved in Section IV. Five healthy male subjects (age = 27 ± 4 years, height = 176 ± 4 cm, and weight = 76 ± 9.8 kg) participated in the experiments. The Doppler radar shown in Fig. 1 and Fig. 6 was used to collect the human data. In Section IV, we showed that using an LP single patch antenna in the Tx side and a CP array in the Rx side results in the best performance among all configurations (the forth row and first column of Fig. 9). Fig. 16 shows the human measurement setup. The subjects wore normal clothing and were seated at a steady position with their chests facing the radar antennas. A wired fingertip pulse sensor from Biopac (MP150-OXE100) was also attached to the index finger of the subjects to provide reference heartbeat signals. All the subjects were asked to stay as motionless as possible. All experiments were approved by the Committee for the Protection of Human Subjects at Stevens Institute of Technology. Fig. 17 shows a sample the frequency spectrum of the 16 different measurement scenarios in a representative human subject. This figure corresponds to the received signal of Subject #2 at the same distance from the target as the actuator (i.e., 1.5 m). The exact same trend and results are observed in the human target as well. This same trend was viewed in all the human subjects which we tested. The best-case scenario is using a single LP antenna at the Tx and a CP array at the Rx.

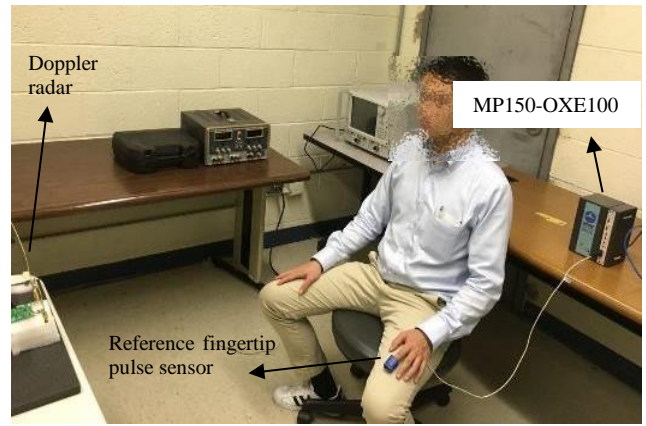

Fig. 16. Measurement setup for heartbeat and respiration detection in human subjects.

To verify and validate this conclusion, we tested and compared the optimum case (LP single patch antenna in the Tx side and a CP array in the Rx ) with the most general case, where Tx and Rx antennas are both LP single patches (the first row and first column in Fig. 9). For each of the two cases, two different distances were considered between the subjects and the radar. First, the distance was only 0.5 m, and therefore the reflected was strong enough and could be recovered easily. Next, the distance was increased to 1.5 m, causing the transmitted signal to experience more losses as well as more reflections from other objects. This led to a degraded quality of the received signal. Data were collected for 5 minutes for each of the four cases from each subject. Although we used a fast Fourier transform (FFT) operation to analyze the linear actuator data, this technique cannot provide an accurate estimation in the case of heartbeat monitoring experiments. Specifically, the frequency resolution of an FFT operation is equal to the inverse of its window size. For a typical window size of 10 seconds, the resolution will be 0.1 Hz, i.e., 6 beats/min. However, to achieve a 2% confidence interval in the case of a typical heartbeat rate of 80 beats/min, a resolution of 0.02 × 80 = 1.6 beats/min would be required. Therefore, the achieved resolution will not be sufficient. A larger window size would provide a better resolution, but would be less capable of tracking rapid rate changes and therefore would not be suitable for real time monitoring. For human experiments, we exploited a well-known method in speech processing named as the "pitch finding technique" [39]. This technique states that: if the signal contains a single dominant frequency, its frequency can be accurately estimated by performing an auto-correlation operation. Therefore, we first applied a fifth order Butterworth band-pass filter with a pass band of 0.85-2.5 Hz (corresponding to a heartbeat rate of 51-150 beats/min) to both I and Q channels. This removes the breathing signal and the higher frequency noise signal. The filtered data was then windowed by a sliding overlapping (100-ms) Hanning window with a duration of 10 seconds. Then, in each window, we took the autocorrelation of (I + jQ). Finally, the FFT of the auto-correlation signal was calculated. Fig. 18 (a) shows a 25-second sample of the raw baseband data from the radar. The distance between the subject and the radar is 0.5 and the best-case antenna configurations are used for Tx and Rx. The spectrum of the data was calculated using the aforementioned method and is plotted in Fig. 18 (b). Note that Fig. 18 (b) shows both the low-pass and band-pass filtered data (respiration and heartbeat signals, respectively). As shown in this figure, the respiration and heartbeat rates for this sample data were correctly found to be about 20 breaths/min and 70 beats/min respectively. Fig. 19 shows the reference heartbeat signal along with the detected heartbeat and respiration signals for the same subject and setup of Fig. 18.

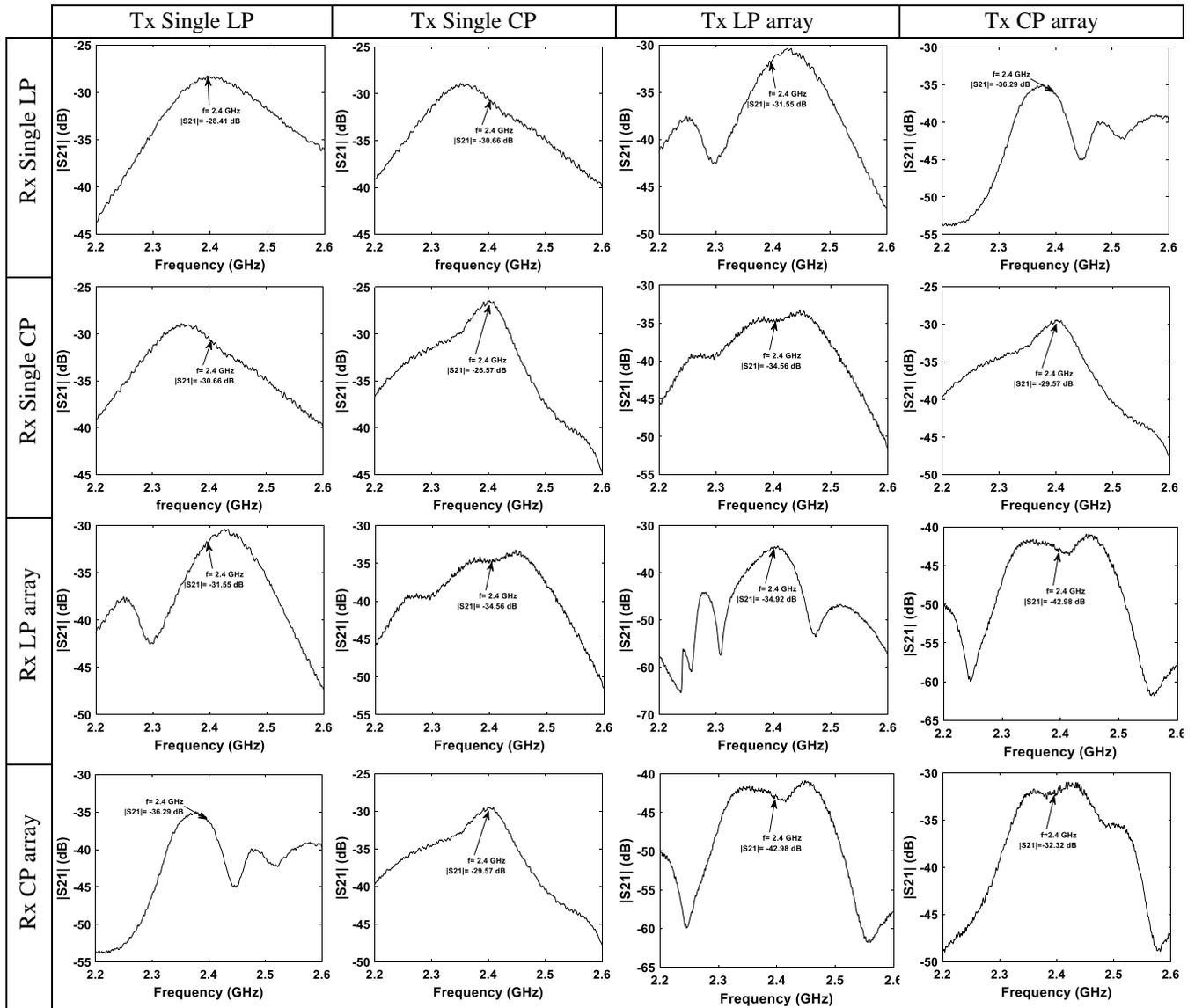

Fig. 17. Frequency spectrum of the different measurement scenarios in a representative human subject. This figure corresponds to the received signal of Subject #2 at the same distance from the target as the actuator (i.e., 1.5 m). The exact same trend and results are observed in the human target as well. The figure is plotted as a 4×4 matrix, which each element of the matrix representing one of the 16 combination sets.

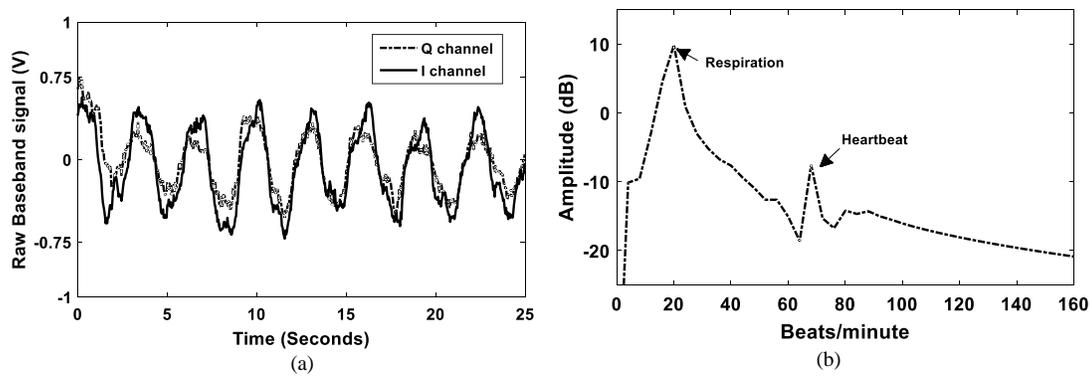

Fig. 18. (a) A 25-second sample of the radar output from a subject with a distance of 0.5 m away from the radar. The best-case configuration was used for the antenna. Both $I$ and $Q$ channel baseband signals are shown. (b) The spectrum of the sampled data. The respiration and heartbeat rates are 20 breaths/min and 70 beats/min respectively.

It is seen that all the peaks in the reference pulse signal are successfully detected in the radar heartbeat signal.

Heart rate detection accuracies were also calculated for each case and are summarized in Table I. The detection accuracy isdefined as the percentage of the time where the detected heartbeat rate from the radar is within 2% of the reference heart rate [6], [10]. It is seen that for a distance of 0.5 m both cases have similar performances. However, when the subjects are 1.5 m away from the radar, the standard case with LP single patch antennas on both Tx and Rx sides has an average of 11% lower detection accuracy compared to the recommended case. The recorded heartbeat signals from Subject 2 while seated 1.5 m away from the radar are depicted in Fig. 20. Fig. 20 (a) demonstrates the standard case where both Tx and Rx use LP single patch antennas while Fig. 20 (b) corresponds to the recommended case where Rx uses a CP array and Tx an LP single patch. The 2% accuracy intervals are also shown. In Fig. 20 (a), 25% of the total measured points (around 75 seconds) are outside of the accuracy interval which implies a detection accuracy of 75%. In contrast, in Fig. 20 (b) only 12% of the measured data (around 36 seconds) are outside of the accuracy interval which shows a detection accuracy of 88%. These results reveal that the recommended case significantly outperforms the standard case in term of detection accuracy.

## VI. CONCLUSION

This paper presents an experimental study on the effects of antenna radiation characteristics on the performance of a Doppler radar used in non-contact heart rate monitoring applications. The effects of the antenna beamwidth and polarization were assessed through a complete set of experiments. Measurement results revealed that careful consideration should be given to the selection of both the Tx and Rx antenna configurations. Experimental results from a linear actuator showed that some specific antenna arrangements outperform other configurations. The results of a proof-of-concept study on human subjects also demonstrated that the suggested antenna configuration has an average of 11% higher detection accuracy compared to the standard case.

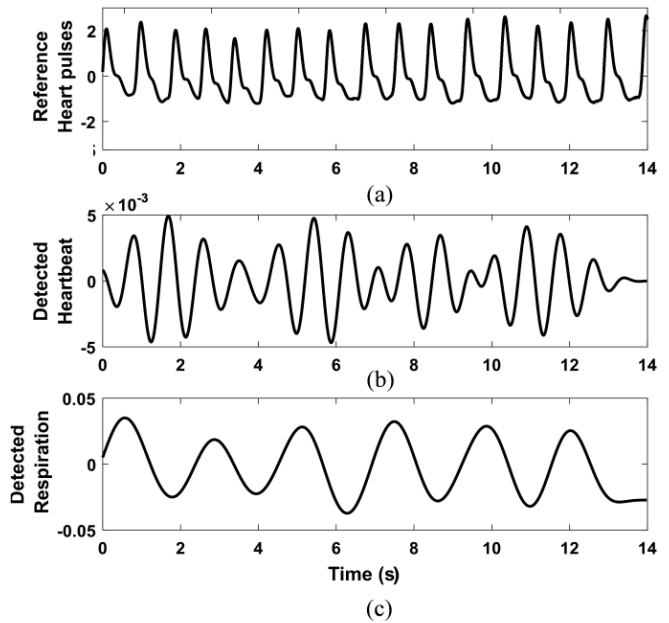

Fig. 19. (a) Reference heartbeat signal, (b) processed radar heartbeat signal, and (c) detected respiration signal from the same subject and setup described in Fig. 18. All the peaks in the reference pulse signal are successfully detected in the radar heartbeat signal.

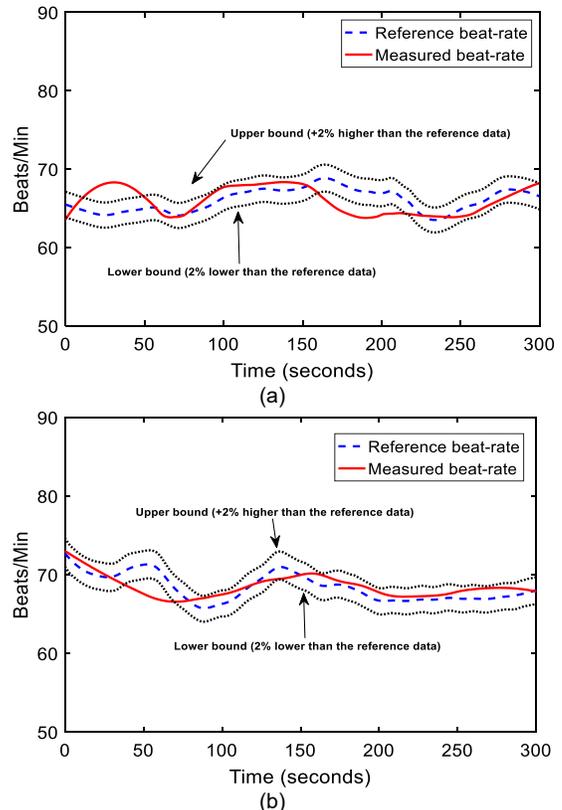

Fig. 20. The recorded heartbeats from Subject 2 while seated 1.5 m away from the radar. The measured data are for a duration of 5 minutes and the heartbeat signals were measured every 10 seconds. (a) Both Tx and Rx use LP single patch antennas. 25% of the total measured points (around 75 seconds) are outside of the accuracy interval which implies a detection accuracy of 75%. (b) Rx uses a CP array and Tx an LP single patch. Only 12% measured points (around 36 seconds) are outside of the accuracy interval which shows a detection accuracy of 88%.

TABLE I
COMPARISON OF THE HEART RATE DETECTION ACCURACIES FOR TWO
DIFFERENT SCENARIOS AT DIFFERENT DISTANCES.

| Radar configuration | Subjects 0.5 m away from the radar | | Subjects 1.5 m away from the radar | |
|---|---|---|---|---|
| | Standard Case | Recommended Case | Standard Case | Recommended Case |
| Subjects | Heart rate accuracy | | Heart rate accuracy | |
| 1 | 99% | 100% | 78% | 89% |
| 2 | 100% | 99% | 75% | 88% |
| 3 | 98% | 100% | 71% | 81% |
| 4 | 98% | 98% | 74% | 85% |
| 5 | 100% | 99% | 77% | 89% |